\documentclass[12pt]{iopart}
\usepackage{color}
\usepackage{iopams} 
\usepackage{graphicx}
\usepackage[colorlinks=true,breaklinks=true] {hyperref}
\usepackage{cite}

\begin{document}
    \title{Bi-Josephson Effect in a Driven-Dissipative Supersolid}
    \author{Jieli Qin$^{1,*}$, Shijie Li$^1$, Yijia Tu$^1$, Maokun Gu$^1$, Lin Guan$^2$, Weimin Xu$^2$, Lu Zhou$^{3,4,\dagger}$}
    
    \address{$^{1}$School of Physics and Materials Science, Guangzhou University, 230
        Wai Huan Xi Road, Guangzhou Higher Education Mega Center, Guangzhou
        510006, China}
    \address{$^{2}$School of Mathematics and Information Science, Guangzhou University, 230
        Wai Huan Xi Road, Guangzhou Higher Education Mega Center, Guangzhou
        510006, China}
    \address{$^{3}$Department of Physics, School of Physics and Electronic Science, East
        China Normal University, Shanghai 200241, China}
    \address{$^{4}$Collaborative Innovation Center of Extreme Optics, Shanxi University,
        Taiyuan, Shanxi 030006, China}
    
    \eads{$^*$\mailto{qinjieli@126.com},$^*$\mailto{104531@gzhu.edu.cn},
        $^\dagger$\mailto{lzhou@phy.ecnu.edu.cn}}
    
    \begin{abstract}
        The Josephson effect is a macroscopic quantum tunneling phenomenon
        in a system with superfluid property, when it is split into two parts
        by a barrier. Here, we examine the Josephson effect in
        a driven-dissipative supersolid
        realized by coupling Bose-Einstein condensates to an optical ring cavity. 
        We show that the spontaneous breaking of spatial translation symmetry in supersolid makes the location of the splitting barrier have a significant influence on the Josephson effect. Remarkably, for the same splitting barrier, depending on its location, two different types of DC Josephson currents are found in the supersolid phase (compared to only one type found in the superfluid phase). Thus, we term it a bi-Josephson effect. We examine
        the Josephson relationships and critical Josephson currents in detail,
        revealing that the emergence of supersolid order affects these two
        types of DC Josephson currents differently---one is enhanced, while
        the other is suppressed. The findings of this work unveil unique Josephson
        physics in the supersolid phase, and show new opportunities to build
        novel Josephson devices with supersolids.
    \end{abstract}
    \noindent{\it Keywords}: Josephson Effect, Supersolid, Bose-Einstein Condensate, Optical Ring Cavity

    \submitto{\NJP}
    
    \section{Introduction \label{sec:introduction}}
    The Josephson effect refers to the macroscopic quantum tunneling phenomena that
    even though matterwaves are split into two parts by a potential barrier,
    a supercurrent can be driven by a phase difference between them \cite{golubov_current-phase_2004,tafuri_fundamentals_2019}.
    It is firstly predicted and experimentally observed in the superconductor
    systems \cite{josephson_possible_1962,nicol_direct_1960,anderson_probable_1963,josephson_discovery_1974},
    and found significant applications in fields such as precision metrology
    \cite{barone_physics_1982,tafuri_fundamentals_2019} and quantum
    computing \cite{barone_physics_1982,tafuri_fundamentals_2019,makhlin_quantum-state_2001,kjaergaard_superconducting_2020,huang_superconducting_2020}.
        Later, the idea extends to many other quantum systems, such as superfluid helium \cite{tafuri_fundamentals_2019,richards_observation_1965,hulin_analog_1972,backhaus_direct_1997,sukhatme_observation_2001}, exciton-polariton condensates \cite{shelykh_josephson_2008,lagoudakis_coherent_2010,abbarchi_macroscopic_2013,munoz_mateo_long_2020,sun_second-order_2021}, optical systems \cite{aihara_optical_1996,ng_precision_2007,ji_josephson_2009,gerace_quantum-optical_2009,fernandez-lorenzo_dissipative_2021}, and also ultracold atomic gases  \cite{williams_nonlinear_1999,zibold_classical_2010,hou_momentum-space_2018,bresolin_oscillating_2023,mukhopadhyay_observation_2024,raghavan_coherent_1999,albiez_direct_2005,levy_c_2007,spagnolli_crossing_2017,burchianti_josephson_2017,xhani_dynamical_2020} which we are concerning in this work. In ultracold atomic gases, the Josephson physics has been extensively studied using spinor condensates \cite{williams_nonlinear_1999,zibold_classical_2010,hou_momentum-space_2018,bresolin_oscillating_2023,mukhopadhyay_observation_2024} and double-well setups \cite{raghavan_coherent_1999,albiez_direct_2005,levy_c_2007,zhang_josephson_2012,spagnolli_crossing_2017,burchianti_josephson_2017,xhani_dynamical_2020}, with abundant new findings including nonlinear Josephson oscillation and self-trapping \cite{williams_nonlinear_1999,raghavan_coherent_1999,albiez_direct_2005}, momentum space Josephson effect \cite{hou_momentum-space_2018,mukhopadhyay_observation_2024}, etc. Recently, the system of homogeneous atomic gases split by a thin barrier which is particularly suitable for examining the DC Josephson effect (for the difference between DC and AC Josephson effect, one may refer to Ref. \cite{levy_c_2007}) also began to attract intense research interests \cite{kwon_strongly_2020,luick_ideal_2020,del_pace_tunneling_2021}.
    
    Supersolid \cite{boninsegni_colloquium_2012}, despite dating back
    to as early as the middle times of the 20th century \cite{gross_unified_1957,thouless_flow_1969,andreev_quantum_1969,chester_speculations_1970,leggett_can_1970},
    is now rising as one of the most active topics in cold atom physics,
    since its successful realization in several different platforms, including
    dipolar Bose-Einstein condensates (BECs) \cite{tanzi_observation_2019,tanzi_supersolid_2019,guo_low-energy_2019,bottcher_transient_2019,chomaz_long-lived_2019,norcia_two-dimensional_2021,sohmen_birth_2021}
    , spin-orbit coupled BECs \cite{li_stripe_2017,bersano_experimental_2019,putra_spatial_2020},
    and optical cavity and BEC coupling systems \cite{leonard_supersolid_2017,leonard_monitoring_2017,mivehvar_driven-dissipative_2018,schuster_supersolid_2020}.
    In this fascinating quantum state, both the phase and spatial translation
    symmetry spontaneously breaks, and matters simultaneously show the superfluid and
    crystalline properties \cite{boninsegni_colloquium_2012}.  It has
    been shown that a DC Josephson current is possible in
    the BEC supersolid induced by finite-range two-body interaction \cite{kunimi_josephson_2011}
    or dipole-dipole interaction \cite{nilsson_tengstrand_toroidal_2023},
    and recently the Josephson effect has been used to measure the superfluid
    fraction of a dipolar supersolid \cite{biagioni_measurement_2024}. 
    
    \begin{figure}
        \begin{centering}
            \includegraphics{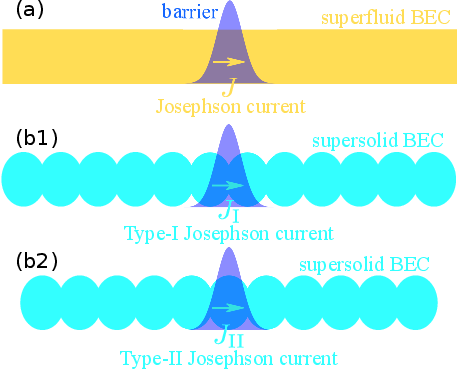}
            \par\end{centering}
        \caption{Illustration of Josephson effect in superfluid (a) vs. supersolid
            (b1, b2). In the spatially uniform superfluid, the position of the barrier do
            not affect the Josephson effect. In the spatially modulated supersolid, depending on the
            location of the barrier {[}at the supersolid valley (b1) or peak (b2){]},
            two different DC Josephson currents would be supported, that is the
            bi-Josephson effect in this work. \label{fig:BiJosephsonEffectIllustration}}
    \end{figure}
    
    In this work, we study the Josephson effect in a driven-dissipative BEC and optical ring
    cavity coupling system \cite{mivehvar_driven-dissipative_2018}, which experiences a superfluid to supersolid phase transition,
    revealing fascinating new characteristics of the Josephson effect
    in the supersolid regime. We illustrate the core physical concept
    in Fig. \ref{fig:BiJosephsonEffectIllustration}. In the Josephson
    effect of typical superfluid, due to the spatial translation symmetry,
    defining a relative position between the superfluid and splitting
    barrier is impossible, thus the Josephson effect would not be affected
    by the location of the barrier. However, for the supersolid, the barrier
    can locate at either the density peak or the density valley of the
    supersolid (and later in the main text we will show that because of
    the parity symmetry protection, only these two cases are allowed),
    this leads to two distinct DC Josephson currents, i.e., the bi-Josephson
    effect in this work. In the following contents, we will elaborate on this concept
    in detail. In Sec. \ref{sec:model}, the system studied in this work and its theoretical description is presented. In Sec. \ref{sec:results}, we show the existence of bi-Josephson effect in supersolid phase, and examine the Josephson relationships and critical Josephson currents. At last, the paper is summarized in Sec. \ref{sec:summary}.
    
    \section{Model \label{sec:model}}
    
    \begin{figure}
    \begin{centering}
        \includegraphics{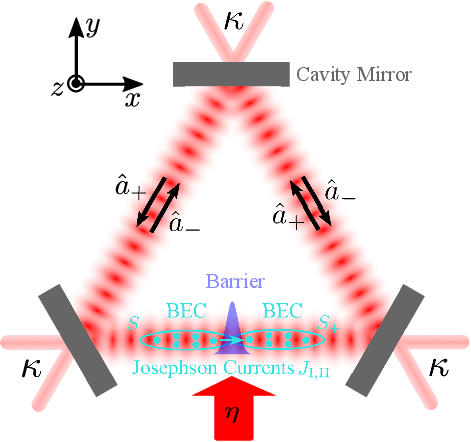}
        \par\end{centering}
    \caption{Schematic diagram of the considering system. Two quasi-one-dimensional
        BECs reside within an optical ring cavity. A laser transversely shining
        on the BECs pumps the system with strength $\eta$. The two counter-propagating
        ring cavity modes ($a_{\pm}e^{\pm ik_{c}x}$) are excited due to the
        two-photon scatterings, and suffer a cavity loss with rate $\kappa$.
        When superradiation takes place, the BECs feel an optical lattice
        potential, and transform from the superfluid state to the supersolid
        state. The two BECs are separated by an external potential barrier,
        a phase difference $\Delta S=S_{+}-S_{-}$ between them can drive
        a DC Josephson current across the barrier.\label{fig:Diagram}}
    \end{figure}

    The system we consider in this work is proposed and realized in Refs.
    \cite{mivehvar_driven-dissipative_2018,schuster_supersolid_2020},
    and we schematically show it in Fig. \ref{fig:Diagram}. Quasi-one-dimensional
    BECs are loaded in a ring cavity along its axis. We pump the
    system by transversely illuminating the BEC atoms using a laser with
    detuning $\Delta_{a}$ and Rabi frequency $\Omega_{0}$. Light fields
    of the two counter-propagating ring cavity modes ($a_{\pm}e^{\pm ik_{c}x}$
    with $a_{\pm}$ being the annihilation operators and $k_{c}$ being
    the wavenumber) are built up as a result of the scattering of pumping
    photons into the cavity. The cavity light fields and BEC atoms interact
    with a strength of ${\cal G}_{0}$. Hamiltonian for this system is
    \cite{mivehvar_driven-dissipative_2018,qin_self-trapped_2020,qin_collision_2021,qin_supersolid_2022}
    \begin{eqnarray}
        H & =-\hbar\Delta_{c}\left(a_{+}^{\dagger}a_{+}+a_{-}^{\dagger}a_{-}\right)+\int\Psi^{\dagger}\left(x\right)H_{a}\Psi\left(x\right)dx\nonumber \\
        & +\frac{1}{2}\int\Psi^{\dagger}\left(x\right)\Psi^{\dagger}\left(x'\right)V\left(x-x'\right)\Psi\left(x'\right)\Psi\left(x\right)dxdx',\label{eq:Hamiltonian_eff}
    \end{eqnarray}
    where the first term accounts for the two counter-propagating cavity
    modes, with $\hbar$ being the Planck constant and $\Delta_{c}$ being
    the detuning between the cavity modes and the pump laser; the last
    term describes the interaction between BEC atoms, with $\Psi$ being
    the BEC field operator, and $V\left(x-x'\right)$ being the interaction
    core. In this work, we consider the repulsive one-dimensional effective
    contact interaction, thus $V\left(x-x'\right)=g_{0}\delta\left(x-x'\right)$
    with $g_{0}>0$ being the interaction strength. The kinetic energy
    of the BEC atoms and the potential felt by them make
    up the second term, with $H_{a}$ being the single particle Hamiltonian
    
    \begin{eqnarray}
        H_{a} & =\frac{p_{x}^{2}}{2m}+V_{\mathrm{total}}\left(x\right).\label{eq:Hamiltonian_atom}
    \end{eqnarray}
    Here, the first term is the kinetic energy, with $m$ being the atomic
    mass and $p_{x}=-i\hbar\frac{\partial}{\partial x}$ being the momentum
    operator. The total potential felt by the atoms consists of two parts,
    $V_{\mathrm{total}}=V_{\mathrm{ext}}+V_{c}$, with $V_{\mathrm{ext}}$
    being an external potential, and $V_{c}$ being the potential due
    to BEC-cavity interaction, which can be further split into two parts,
    $V_{c}=V_{c,1}+V_{c,2}$. The two-photon scattering between the two
    cavity modes leads to
    \begin{equation}
        V_{c,2}=\hbar U_{0}\left[a_{+}^{\dagger}a_{+}+a_{-}^{\dagger}a_{-}+\left(a_{+}^{\dagger}a_{-}e^{-2ik_{c}x}+\mathrm{h.c.}\right)\right],\label{eq:V_ac}
    \end{equation}
    and the two-photon scattering between the pump laser and the cavity
    modes leads to
    \begin{equation}
        V_{c,1}=\hbar\eta_{0}\left(a_{+}e^{ik_{c}x}+a_{-}e^{-ik_{c}x}+\mathrm{h.c.}\right).\label{eq:V_ap}
    \end{equation}
    Their strengths are respectively $\hbar U_{0}=\hbar\mathcal{G}_{0}^{2}/\Delta_{a}$ and
    $\hbar\eta_{0}=\hbar\mathcal{G}_{0}\Omega_{0}/\Delta_{a}$. In the
    following contents, we would apply the natural units $m=\hbar=k_{c}=1$
    for simplicity of the formulae.
    
    Applying the mean field theory \cite{mivehvar_cavity_2021} (we note that although mean field theory may miss quantum characteristics such as correlations and fragmentation \cite{lode_fragmented_2017}, it is accurate in the thermodynamic limit \cite{piazza_boseeinstein_2013,li_stochastic_2024}, and can capture the main supersolid physics in the present system, with the mean field results well fitting the experimental observations \cite{schuster_supersolid_2020}),
    the quantum operators $a_{\pm}$ and $\Psi$ are approximately replaced
    by their mean values, i.e., $a_{\pm}/\sqrt{N}\rightarrow\alpha_{\pm}$,
    $\Psi/\sqrt{N}\rightarrow\psi$ {[}here we also scale them by the
    total atom number $N$, such that $\psi$ is normalized to one, i.e., $\int\left|\psi\left(x\right)\right|^{2}dx=1${]}.
    Taking the mean values of the corresponding Heisenberg equations,
    $\alpha_{\pm}$ and $\psi$ are governed by
    \begin{equation}
        i\frac{\partial}{\partial t}\alpha_{\pm}=\left(-\Delta_{c}+U-i\kappa\right)\alpha_{\pm}+UN_{\pm2}\alpha_{\mp}+\eta N_{\pm1},\label{eq:Meanfield_cavity}
    \end{equation}
    \begin{equation}
        i\frac{\partial}{\partial t}\psi=\left[-\frac{1}{2}\frac{\partial^{2}}{\partial x^{2}}+V_{\mathrm{total}}\left(x\right)\right]\psi+g\left|\psi\right|^{2}\psi,\label{eq:Meanfield_atom}
    \end{equation}
    where
    \begin{equation}N_{\pm m}=\int\left|\psi\left(x\right)\right|^{2}e^{\mp imx}dx, \label{eq:AtomOrderParameters}
    \end{equation}
    with $m=1,2$ are atomic order parameters characterize the probability
    of two-photon scatterings ($N_{\pm1}$ for two-photon scattering between
    the pump laser and the cavity modes, while $N_{\pm2}$ for two-photon
    scattering between the two cavity modes); $U=NU_{0}$, $\eta=\sqrt{N}\eta_{0}$,
    $g=Ng_{0}$ are the scaled two-photon scattering and contact interaction
    strengths; and here the cavity-photon loss with rate $\kappa$ is
    introduced phenomenological. 
    
    Due to the balance between the pumping and cavity loss, the system
    will reach a steady state, which can be mathematically obtained by
    letting $\partial_{t}\alpha_{\pm}=0$, and $\psi\left(x,t\right)=\psi\left(x\right)e^{-i\mu t}$
    with $\mu$ being the chemical potential. Inserting them into equations
    (\ref{eq:Meanfield_cavity}) and (\ref{eq:Meanfield_atom}), one gets
    the following time-independent equations for steady state
    
    \begin{equation}
        \mu\psi=\left[-\frac{1}{2}\frac{\partial^{2}}{\partial x^{2}}+V_{\mathrm{total}}\left(x\right)\right]\psi+g\left|\psi\right|^{2}\psi,\label{eq:steadypsi}
    \end{equation}
    \begin{equation}
        \alpha_{\pm}=-\frac{\left(-\Delta_{c}+U-i\kappa\right)\eta N_{\pm1}-\eta UN_{\pm2}N_{\mp1}}{\left(-\Delta_{c}+U-i\kappa\right)^{2}-U^{2}N_{-2}N_{+2}}.\label{eq:alpha_pm}
    \end{equation}

    To examine the Josephson effect, we separate the BECs with
    an external potential barrier located at $x=0$, 
    \begin{equation}
        V_{\mathrm{ext}}\left(x\right)=V_{0}\exp\left[-\left(\frac{x}{\sigma}\right)^{2}\right],\label{eq:extPotential}
    \end{equation}
    where $V_{0}$ and $\sigma$ are the height and width of the barrier.
    According to the Josephson effect physics, a phase jump of the wavefunction
    $\psi$ across the barrier would drive a constant current $J=-i\left(\psi^{*}\partial_{x}\psi-\psi\partial_{x}\psi^{*}\right)/2$
    in the system. To better theoretically understand this, we rewrite
    the wavefunction as $\psi\left(x\right)=A\left(x\right)\exp\left[iS\left(x\right)\right]$,
    with real functions $A\left(x\right)$ and $S\left(x\right)$ being
    its amplitude and phase distribution. In terms of $A$ and $S$, the
    current $J$ can be rewritten as $J=A^{2}\left(dS/dx\right)$. Inserting
    $\psi=A\exp\left(iS\right)$ into Eq. (\ref{eq:steadypsi}), and splitting
    the real and imaginary parts, we firstly obtain that $dJ/dx=0$, which
    means that in a steady state, only a spatially uniform current can
    exist; and we also get another equation describes the amplitude of
    the wavefunction
    \begin{equation}
        \mu A=\left(-\frac{1}{2}\frac{d^{2}}{dx^{2}}+\frac{1}{2}\frac{J^{2}}{A^{4}}+V_{\mathrm{total}}+gA^{2}\right)A.\label{eq:staintionaryA}
    \end{equation}
    For a given DC Josephson current $J$, we solve Eq. (\ref{eq:staintionaryA})
    {[}together with Eq. (\ref{eq:alpha_pm}){]} in the range of $x\in\left[-L/2,L/2\right]$
    (numerically we choose $L=40\pi$ in this work), under the boundary
    condition $A\left(x-L/2\right)=A\left(x+L/2\right)$. The one-dimensional
    system with such a boundary condition can be regarded as an orifice
    attaching two reservoirs of superfluid or supersolid \cite{anderson_considerations_1966,kunimi_josephson_2011}.
    After $A\left(x\right)$ is solved, the phase distribution can be
    obtained by integrating $J=A^{2}\left(dS/dx\right)$, that is $S\left(x\right)=J\int_{0}^{x}1/A^{2}\left(\xi\right)d\xi$
    {[}in this form, the phase at $x=0$ is fixed to zero, $S\left(x=0\right)=0${]}.
    
    \begin{figure}
        \begin{centering}
            \includegraphics[clip]{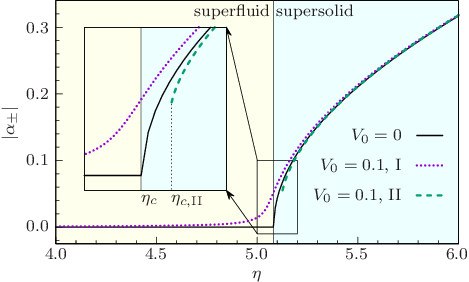}
            \par\end{centering}
        \caption{Superradiant phase transition. The cavity field amplitude $\left|\alpha_{\pm}\right|$
            is plotted as a function of the pumping strength $\eta$. The no external
            potential ($V_{0}=0$) case is plotted with the solid black line.
            The violet dotted and green dashed lines correspond to the type I
            and II solutions under external potential $V_{0}=0.1$. The inset
            shows an enlargement of the lines in the rectangle area. For all the
            lines, other parameters are $U=-0.5,$ $\Delta_{c}=-1$, $\kappa=10$,
            $g=1$, $\sigma=1$ (these parameters will keep fixed all through
            this work) and $J=0.8J_{0}$ with $J_{0}=N\hbar/\left(mL^{2}\right)$.
            \label{fig:PhaseTransition}}
    \end{figure}
    
    \section{Results \label{sec:results}}
    Before diving into the main results of this work, let us first have a brief review on the superradiant (or superfluid to supersolid) phase transition in the system when there is no external barrier ($V_{\mathrm{ext}}=0$) \cite{mivehvar_driven-dissipative_2018}. In this case, the system
    exhibits a continuous $U\left(1\right)$ symmetry. It is invariant
    under spatial translation $x\rightarrow x+X$ and cavity phase rotations
    $a_{\pm}\rightarrow a_{\pm}e^{\mp ik_{c}X}$. Under weak pumping (small
    $\eta$), the cavity modes are practically empty. However, as the
    pumping strength $\eta$ increases beyond a critical value $\eta_{c}$,
    a superradiant phase transition happens. The cavity fields $\alpha_{\pm}$
    are quickly built up with their relative phase fixed to an arbitrary
    value between 0 and $2\pi$. This scenario is shown in Fig. \ref{fig:PhaseTransition}
    with the black solid line. From the perspective of BEC, below $\eta_{c}$,
    the absence of cavity field lattice potential allows the condensate
    to have a uniform density; however, above $\eta_{c}$, the building
    up of the superradiant cavity lattice potential spontaneously breaks the continuous
    spatial translation symmetry, and forces the condensate density to
    adopt a periodical modulation, signifying a superfluid to supersolid
    phase transition.
    Here, We emphasize that the superradiant cavity lattice has a substantial difference from an externally applied optical lattice: Since the cavity light field is built up by pumping the BEC, the phases of the two optical modes are determined by the quantum state of the condensate, thus the superradiant optical lattice resulting from the interference of these two optical modes can automatically adjust its location with respect to the condensate \cite{mivehvar_driven-dissipative_2018,gietka_supersolid-based_2019,qin_self-trapped_2020,qin_collision_2021,qin_supersolid_2022}.
    
    \begin{figure}
        \begin{centering}
            \includegraphics{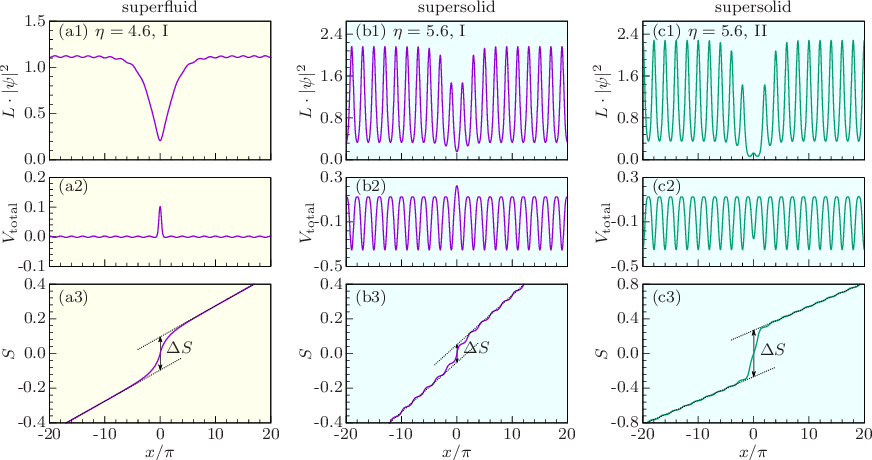}
            \par\end{centering}
        \caption{Bi-Josephson effect. The BEC density $\left|\psi\left(x\right)\right|^{2}=A\left(x\right)^{2}$,
            total potential $V_{\mathrm{total}}\left(x\right)$, and phase distribution
            $S\left(x\right)$ are respectively plotted in the upper (a1-c1),
            middle (a2-c2) and bottom (a3-c3) panels. In the left panels (a1-a3),
            the pumping strength is $\eta=4.6$ (superfluid phase). While in the
            middle (b1-b3) and right (c1-c3) panels, it is $\eta=5.6$ (supersolid
            phase). The solutions shown in the left (a1-a3) and middle (b1-b3)
            panels belong to type I (around $x=0$, the total potential is a local
            maximum, and the atomic density is a local minimum). The solution
            shown in the right panels (c1-c3) belongs to type II (around $x=0$,
            the total potential is a local minimum, and the atomic density is
            a local maximum even though very small). In panels (a3-c3), the black
            dashed lines are linear fits of the curves in the region away from
            the external barrier. The phase jump $\Delta S$ is estimated using
            these linear fits, and they are 0.19 (a3), 0.10 (b3) and 0.53 (c3),
            respectively. For all the panels, the external barrier height is $V_{0}=0.1$,
            and the Josephson current is $J=0.8J_{0}$. \label{fig:DCJosephson}}
    \end{figure}
    
    Introducing the external potential barrier (\ref{eq:extPotential}) into the system
    explicitly breaks the continuous spatial translation symmetry, leading to a nonuniform BEC density. Recalling the atomic order parameters definition Eq. (\ref{eq:AtomOrderParameters}),
    one easily concludes that this nonuniform BEC density would usually give rise to non-zero values of $N_{\pm1,2}$. Then, according to Eq. (\ref{eq:alpha_pm}), this will consequently result in
    a non-empty cavity light field, even below the original superradiant
    critical value $\eta_{c}$. In essence, the sharp superradiant phase
    transition is blurred by the barrier, as shown by the violet dotted line in
    Fig. \ref{fig:PhaseTransition}. 
    
    For the superfluid BEC,
    it feels a total potential, which is a combination of the weak cavity
    field lattice and the external barrier, as shown in Fig. \ref{fig:DCJosephson}(a2). Therefore,
    the external barrier drives the atoms away, creating a density dip around $x=0$, and the weak cavity field lattice imprints a slight periodical modulation on the BEC density [Fig. \ref{fig:DCJosephson}(a1)]. Since we impose a constant current
    in the condensate, the phase has a linear distribution away from the
    external barrier region (the two black dotted lines are linear fits
    of the numerical results $S_{\pm}=\beta_{\pm}x+S_{0,\pm}$ with $\pm$
    for the right and left sides respectively); and across the barrier, the
    phase experiences a jump, which we estimate as $\Delta S=S_{0,+,}-S_{0,-}$, see Fig. \ref{fig:DCJosephson}(a3).
    This phase jump and the consequent current flow through the barrier signify the Josephson effect.
    
    In the supersolid phase, under the same system parameters, two different
    types of steady state solutions are found, which we will refer to
    as type I and II in the following contents. For the type I solution,
    the cavity field lattice
        $V_c=V_1\cos\left(k_c x + \phi_1\right) + V_2 \cos\left(2k_c x+\phi_2\right)$ [here $V_{1,2},\phi_{1,2}$ are determined by the BEC wavefunction through Eqs. (\ref{eq:V_ac},\ref{eq:V_ap},\ref{eq:AtomOrderParameters},\ref{eq:alpha_pm})] and the external barrier Eq. (\ref{eq:extPotential}) are in
        phase (i.e., $\phi_{1,2}=0$, such that the cavity lattice barrier meets the external barrier),
    and the superposition of them arises a higher bump at $x=0$ {[}Fig.
    \ref{fig:DCJosephson}(b2){]}. Under such a total potential, the supersolid
    BEC profile exhibits an overall dip structure around the external
    barrier, and exactly at $x=0$, the BEC density is a local minimum
    {[}Fig. \ref{fig:DCJosephson}(b1){]}. Having a review of the superfluid
    phase solution shown in Fig. \ref{fig:DCJosephson}(a1-a3), one immediately
    finds that it also belongs to this category. For the type II solution,
    the cavity field lattice and the external barrier are in
    opposite phases (i.e., 
    $\phi_{1,2}=\pi$,
    such that the cavity lattice well meets the external
    barrier). The superposition of them yields a total potential shown
    in Fig. \ref{fig:DCJosephson}(c2), where at $x=0$ it is a local
    potential well shallower than other lattice sites. Under this type
    of potential, the supersolid BEC profile still shows a overall dip
    structure around the external barrier. However, exactly at $x=0$,
    the BEC density can be a local maximum, albeit much smaller than other
    supersolid peaks, see Fig. \ref{fig:DCJosephson}(c1). As the pumping
    strength $\eta$ weakens, the depth of the local well at $x=0$ decreases.
    Eventually, below a critical value $\eta_{c,\mathrm{II}}$, this type
    of solution can no longer be supported, this is illustrated by the
    green dashed line in Fig. \ref{fig:PhaseTransition}, which stops at $\eta_{c,\mathrm{II}}$. 
    No solutions were found, where the induced cavity field lattice and the external barrier have relative phase $\phi_{1,2}\neq 0,\pi$. 
    This would be understood as follows: The system under consideration owns a parity symmetry, i.e., Eq. (\ref{eq:staintionaryA}) is parity symmetric under translation $x\rightarrow -x$, $\alpha_\pm \rightarrow \alpha_\mp$. Therefore, the solutions also deserve a corresponding symmetry. However, superposition
    of the induced cavity field lattice and external barrier with other relative
    phases $\phi_{1,2}\neq 0,\pi$ would violate this symmetry, which makes such configurations not allowed. The superradiant cavity field lattice will always automatically adjust its location to line up its maximum or minimum with the external barrier.

    \begin{figure}
    \begin{centering}
        \includegraphics{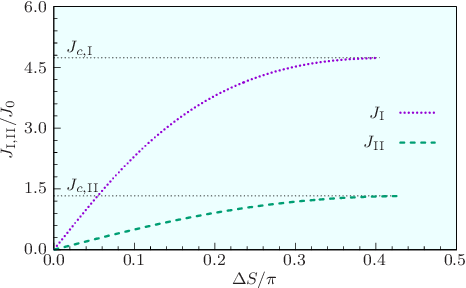}
        \par\end{centering}
    \caption{Bi-Josephson relationships $J_{\mathrm{I}}$-$\Delta S$ (violet solid
        line) and $J_{\mathrm{II}}$-$\Delta S$ (green dashed line). The
        horizontal black dotted lines illustrate the maximum current, i.e.,
        the critical Josephson currents $J_{c,\mathrm{I}}$ and $J_{c,\mathrm{II}}$.
        The parameters used are the same as those in Fig.\ref{fig:DCJosephson}
        (b1-b3, c1-c3), except that $J$ is varying. \label{fig:JosephsonRelation}}
\end{figure}
    
    In Fig. \ref{fig:DCJosephson}(b3,c3), we see that the phase jumps across the barrier are quite different, although the two types of solutions carry the same DC Josephson current.
    We further examined the Josephson relationship of the system, and
    two distinctive $J$-$\Delta S$ relationships are found corresponding
    to the two solution types, see Fig. \ref{fig:JosephsonRelation}.
    In this sense, we term it the bi-Josephson effect. Both the two Josephson
    relationships have their own critical current ($J_{c;\mathrm{I,}}$
    and $J_{c,\mathrm{II}}$ illustrated by the two horizontal black dotted
    lines in Fig. \ref{fig:JosephsonRelation}), above which steady state
    solutions can no longer be found. 

\begin{figure}
        \begin{centering}
            \includegraphics{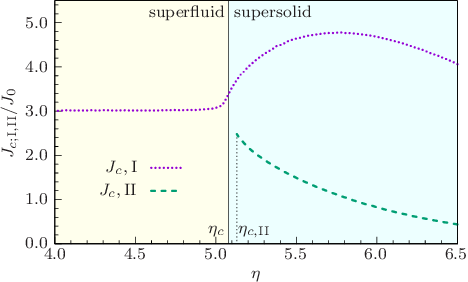}
            \par\end{centering}
        \caption{Critical bi-Josephson currents $J_{c,\mathrm{I}}$ (violet dotted
            line) and $J_{c,\mathrm{II}}$ (green dashed line) as a function of
            pumping strength $\eta$. The parameters used are the same as those
            in Fig.\ref{fig:JosephsonRelation}, except that $\eta$ is varying.
            \label{fig:CriticalCurrents}}
    \end{figure}

    The change of the critical Josephson currents
    during the superfluid to supersolid phase transition is shown in Fig.
    \ref{fig:CriticalCurrents}. Before the superradiant phase transition
    takes place, increasing the pumping strength $\eta$ has little effect
    on the system. The cavity field remains almost empty, so the critical
    Josephson current in the superfluid BEC is barely affected, see the
    plateau in the left part of Fig. \ref{fig:CriticalCurrents}. After
    the superradiant phase transition, the type I critical Josephson current
    $J_{c,\mathrm{I}}$ firstly increases with the pumping strength $\eta$.
    This increasing can be understood by comparing panels (a1) and (b1)
    of Fig. \ref{fig:DCJosephson}. In these two panels, the external
    potential barrier is the same, but compared to the superfluid in panel
    (a1), the left-to-right separation of the supersolid in panel (b1)
    is much less obvious. This indicates that the emergence of supersolid
    order can reduce the splitting effect of the external barrier, and enhance the type I
    supercurrent. Further increasing of $\eta$ leads to a deep superradiant
    optical lattice, which will break the BEC into a series of almost
    isolated droplets. This weakens the supercurrent, causing $J_{c,\mathrm{I}}$
    to decrease. For the type II critical Josephson current, by comparing
    panels (a1) and (c1) of Fig. \ref{fig:DCJosephson}, we conclude that
    the emergence of supersolid order would enhance the splitting effect of the
    external barrier, and suppress the type II supercurrent. Therefore, as $\eta$
    increases, $J_{c,\mathrm{II}}$ always decreases. 
    These observations suggest that the Josephson physics in supersolid would be engineered more flexible, and new types of Josephson devices might be constructed (for example, one may encode information into the two different Josephson currents).
    
    \section{Summary \label{sec:summary}}
    In summary, we reveal an essential new Josephson phenomenon in supersolid, which is absent in the usual superfluid. We show that in supersolid due to the spontaneous spatial translation symmetry breaking, depending on the location of the splitting barrier, two different types of Josephson currents can be supported, in sharp contrast
    with the usual superfluid, in which only one Josephson current is
    observed. We call this phenomenon the bi-Josephson effect. We examine this effect in detail, and give the Josephson relationships and critical Josephson currents.   
 
    In this work, we demonstrate the bi-Josephson effect in a driven-dissipative supersolid achieved
    through BEC and ring-cavity coupling. We expect the bi-Josephson effect to be a general
    phenomenon in supersolids across various platforms \cite{tanzi_observation_2019,tanzi_supersolid_2019,guo_low-energy_2019,bottcher_transient_2019,chomaz_long-lived_2019,norcia_two-dimensional_2021,sohmen_birth_2021,li_stripe_2017,bersano_experimental_2019,putra_spatial_2020,leonard_supersolid_2017,leonard_monitoring_2017},
    and exploring this in future studies would be of great interest. We
    hope the bi-Josephson physics in supersolid would contribute to
    the development of novel Josephson devices with applications in the
    precision measurement and quantum information fields.
    
    \section*{Acknowledgments}
        This work is supported by Guangdong Basic and Applied Basic Research
        Foundation (2024A1515012526), National Natural Science Foundation
        of China (11904063, 12074120, 11374003), Innovation Program of the Shanghai Municipal Education Commission (Grant No. 202101070008E00099) and Shanghai Science and Technology Innovation Project (No. 24LZ1400600) .
        
    \section*{Reference}
\providecommand{\newblock}{}
\providecommand{\url}[1]{{\tt #1}}
\providecommand{\urlprefix}{}
\providecommand{\href}[2]{#2}

\end{document}